# Temperature Hysteresis of Magnetization in Lanthanum Manganite

A. P. Saiko* and S. A. Markevich

*Institute of Solid State and Semiconductor Physics, National Academy of Sciences of Belarus, Minsk, 220072 Belarus*

*\* e-mail: saiko@ifttp.bas-net.by*

**Abstract**—We propose an explanation for the experimentally observed temperature hysteresis of magnetization in single crystals of lanthanum manganite ($La_{0.8}Sr_{0.2}MnO_3$). The phenomenon is interpreted within the framework of a double-exchange model with allowance for the interaction of the magnetic subsystem with a bistable mode of the tilting-rotational oscillations in the correlated sublattice of $MnO_6$ octahedra.

As is known, strong interactions between the magnetic, electron, and lattice subsystems in lanthanum manganites $La_{1-x}D_xMnO_3$ (D = Ca, Sr, Ba) account for the large variety of features observed in the physical properties of these compounds, which exhibit structural transformations, magnetic transitions, concentration- and temperature-induced metal–dielectric transitions, etc.

This paper is devoted to the temperature hysteresis of magnetization in single crystals of $La_{1-x}Sr_xMnO_3$ (LSMO). These lanthanum manganites possess orthorhombic ($P_{nma}$) lattices at low temperatures and exhibit a transition (with respect to both concentration and temperature) to a rhombohedral ($R\bar{3}c$) phase in the vicinity of room temperature (the compositions with $0.2 < x < 0.7$ remain orthorhombic at all temperatures).

The structural phase transitions both between $P_{nma}$ and $R\bar{3}c$ phases and between different modifications (O', O*) of the orthorhombic phase of LSMO were thoroughly studied using the measurements of sound velocity and internal friction in single crystals. The discovery of the temperature hysteresis of the sound velocity in $La_{0.8}Sr_{0.2}MnO_3$ led to a conclusion concerning the possible coexistence of orthorhombic and rhombohedral phases in a broad temperature range [1]. Based on these results, we have previously developed a theoretical description [2] of the bistable behavior of elastic characteristics of LSMO crystals.

It is interesting to note that a temperature hysteresis (in the interval from 80 to 110 K) is also characteristic of the behavior of magnetization in $La_{0.8}Sr_{0.2}MnO_3$ samples. The width of the magnetic hysteresis loop along the temperature axis is independent of the applied magnetic field, but the amplitude of this hysteresis (i.e., the difference of magnetizations on cooling and heating in the region of bistability) decreases with increasing magnetic field.

The phenomenon of hysteresis in the magnetization can be explained within the framework of the so-called double-exchange model with allowance for the interaction of the magnetic subsystem with lattice oscillations, in particular, with the bistable mode of the tilting-rotational oscillations of $MnO_6$ octahedra [2]. According to the double-exchange model, three $3d$ electrons of manganese ion (populating the $t_{2g}$ orbital) form a localized spin due to the Hund interaction $J_H$ with mobile electrons on the $e_g$ orbitals of $Mn^{3+}$. This interaction gives the main contribution to the electric conductivity and the ferromagnetic interaction between manganese ions in $La_{0.8}Sr_{0.2}MnO_3$ crystals. The double exchange Hamiltonian can be written as

$$H = -t \sum_{\langle i,j \rangle, \sigma} (c_{i\sigma}^+ c_{j\sigma} + \text{h.c.}) - J_H \sum_i \mathbf{S}_i \boldsymbol{\sigma}_i \\ - \sum_i g\mu_B B S_i^z, \quad (1)$$

where $c_{i\sigma}^+$ is the operator of $e_g$ electron production with spin $\sigma$ at the $\mathbf{R}_i$ site and $t$ is the energy of $e_g$ electron hopping between adjacent sites. The third term in Hamiltonian (1) represents the Zeeman energy of localized spins $S_i$ in the magnetic field $B$.

In the limit as $J_H \longrightarrow \infty$, Hamiltonian (1) reduces to an effective Hamiltonian with independent charge and

spin excitations:

$$H_{\text{eff}} = H_\rho + H_\sigma, \quad (2)$$

$$H_\rho = -t\sum_{\langle j,j\rangle} f_i^+ f_j, \quad H_\sigma = -J_{DE}\sum_{\langle i,j\rangle} \mathbf{S}_i \mathbf{S}_j, \quad (3)$$

where $f_i^+$ is the operator of electron production with a spin directed along the local spin vector $\mathbf{S}_i$. The constant of ferromagnetic interaction depends on the kinetic energy of itinerant (band) electrons averaged over the ground state:

$$J_{DE} = \frac{t}{4S^2}\sum_{\langle i,j\rangle}\langle f_i^+ f_j\rangle_0. \quad (4)$$

The charge part of Hamiltonian (2) describes band electrons with the dispersion $\varepsilon_k = -zt\gamma_k$, while the spin part describes spin-wave excitations with the dispersion $\omega_k = zJ_{DE}S(1-\gamma_k)$, where $z$ is the number of nearest neighbors, $S$ is the total spin, and $\gamma_k = (\cos k_x + \cos k_y + \cos k_z)/3$ is the structural factor for a cubic lattice of Mn sites in units of the lattice constant $a = 1$.

Apparently, the spin–spin ferromagnetic interaction (3) must also depend in the general case on the Mn–O–Mn bond angle, which increases with amplitude of the tilting-rotational oscillations of $MnO_6$ octahedra. This circumstance can be taken into account by introducing an additional interaction term into the Hamiltonian (2),

$$H_\sigma^\xi = -\xi q^2 \sum_{\langle i,j\rangle}\mathbf{S}_i\mathbf{S}_j, \quad (5)$$

which represents a relation between the magnetic subsystem and the correlated subsystem of $MnO_6$ octahedra oscillating in the bistable potential $V(q) + \alpha q^2/2 - \beta q^3/3 + \gamma q^4/4$ [2], where $\xi$ is the coupling coefficient; $\alpha$, $\beta$, $\gamma$ are positive constants; and $q$ is the generalized coordinate describing the amplitude of oscillations. The quadratic dependence on the generalized coordinate expresses the symmetry of term (5) with respect to the reversal of time and configuration space.

In the mean field approximation, the effective temperature-renormalized constant of ferromagnetic interaction can be written as

$$J_{DE}^{\text{eff}} = J_{DE} + \xi\langle q^2\rangle. \quad (6)$$

Nonlinear oscillations of the bistable sublattice will be considered within the framework of the method of self-consistent phonons. For simplicity, let us ignore the reverse influence of the magnetic subsystem on this sublattice. Then, using the spin-wave approximation,

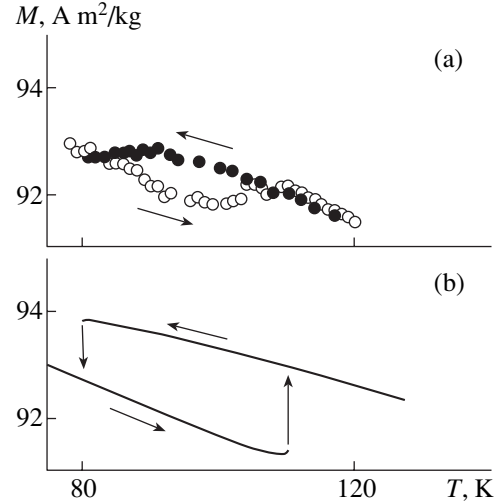

**Fig. 1.** Temperature dependence of the magnetization $M$ of $La_{0.8}Sr_{0.2}MnO_3$ single crystals in the vicinity of a phase transition: (a) experiment (open and black circles correspond to heating and cooling, respectively); (b) theoretical curve calculated for $a = 5\times 10^{-5}$ K$^{-3/2}$; $\xi/J_{DE} = 10^{18}$, $g = 2$, $q_1 = 0.073$ Å, $q_2 = 0.132$ Å, $U_0 = 0.014$ eV, and $m = 96$ amu.

we eventually obtain the following expression for the magnetization:

$$M(T) \approx M_0\left(1 - \frac{aT^{3/2}\exp(-\beta\Delta)}{\left[1 + \dfrac{\xi}{J_{DE}}(\sigma + \langle q\rangle^2)\right]^{3/2}}\right), \quad (7)$$

where $M_0 = M(0)$, $\beta = 1/k_B T$, $\Delta = g\mu_B H$, $g$ is the Lande factor, $\mu_B$ is the Bohr magneton, $H$ is the magnetic field, $\zeta$ is the Riemann function, $a$ is a parameter, $\langle q\rangle$ is the average amplitude of oscillations, and $\sigma = \langle q^2\rangle - \langle q\rangle^2$ is the dispersion. The later two quantities are self-consistently determined from the following system of equations:

$$(\beta - 3\gamma\langle q\rangle)\sigma = \alpha\langle q\rangle - \beta\langle q\rangle^2 + \gamma\langle q\rangle^3,$$

$$\sigma = \frac{1}{2m\Omega}\coth\frac{\Omega}{2k_B T},$$

$$m\Omega^2 = \alpha - 2\beta\langle q\rangle + 3\gamma(\sigma + \langle q\rangle^2),$$

where $m$ is the mass of the oxygen octahedron; $\alpha$, $\beta$, $\gamma$ are constants defined via the physical parameters of the asymmetric two-well potential as

$$\alpha = \frac{12q_2 U_0}{q_1^2(2q_2 - q_1)}, \quad \beta = \frac{12U_0(q_1 + q_2)}{q_1^3(2q_2 - q_1)},$$

$$\gamma = \frac{12U_0}{q_1^3(2q_2 - q_1)},$$

$q_1$ and $q_2$ are generalized coordinates of the potential barrier and metastable minimum, respectively; and $U_0$ is the barrier height in the energy units (see [2]).

The experimentally measured and theoretical temperature dependences of the magnetization of lanthanum manganite $La_{0.8}Sr_{0.2}MnO_3$ are presented in Fig. 1 (the values of $\alpha$, $\beta$, $\gamma$, and $m$ used in the calculations are presented in [2]).

At low temperatures, the sublattice of $MnO_6$ octahedra oscillates in the global minimum of the bistable potential. As the crystal is heated, the system jumps to the state of over-barrier oscillations; the reverse transition (on cooling) to oscillations in the global potential minimum proceeds with a certain delay (i.e., the so-called "supercooled" states are realized, see [2, 3]). This temperature hysteresis of the correlated sublattice of $MnO_6$ octahedra is manifested in the temperature variation of the effective constant of ferromagnetic interaction (6), which depends on the bistable fluctuations of the mean oscillation amplitude $\langle q^2 \rangle$. This relationship leads eventually to the temperature hysteresis in the magnetization of $La_{0.8}Sr_{0.2}MnO_3$ crystals observed in the interval from 80 to 110 K. The experimental data agree both with the theoretical temperature interval of bistability and with the direction of loop traversing: the branch of the magnetization curve measured on heating is below that measured on cooling. An analogous behavior (in the temperature interval from 190 to 210 K) was observed for $La_{0.8}Ba_{0.2}MnO_3$ single crystals [4].

In concluding, it should be noted that $La_{0.8}Sr_{0.2}MnO_3$ single crystals also exhibit a hysteresis of the ultrasound wave velocity, damping, and specific heat capacity [1], while $La_{0.8}Ba_{0.2}MnO_3$ single crystals also show a hysteresis of electric resistance [4].